\documentclass[twocolumn,showpacs,superscriptaddress,amsmath,amssymb,aps,pra]{revtex4-2}

\usepackage{graphicx}
\usepackage{CJKutf8}
\usepackage{dcolumn}
\usepackage{amsmath,bm}
\usepackage{epstopdf}
\usepackage[mathlines]{lineno}
\usepackage{color}
\usepackage[colorlinks=true,linkcolor=blue,citecolor=magenta,urlcolor=cyan]{hyperref}

\begin{document}
\title{Topology of Wilson-loop spectrum and periodic evolution of surface-state Fermi arc}
\author{Yi-Dong Wu}
\email{wuyidong@ysu.edu.cn}
\affiliation{Key Laboratory for Microstructural Material Physics of Hebei Province, School of Science, Yanshan University, Qinhuangdao 066004, China}

\begin{abstract}
 Wilson-loop has been widely used to characterize the topological property of topological insulators, high order topological insulators and topological semimetals. Both bulk topological invariants and nontrivial boundary properties can be deduced from the topology of Wilson-loop spectrum. However, no attempt has been made to observe it. In this letter we demonstrate the topology of Wilson-loop spectrum can be observed by using the existing technology for observing surface-state Fermi arc. We predict that the surface-state Fermi arc sweeps the whole surface Brillouin zone in a continuous periodic process and thus generates a surface that is topologically equivalent to the Wilson-loop spectrum. So by observing the evolution of surface-state Fermi arc in the periodic process we can get the topology of Wilson-loop spectrum.
\end{abstract}
\date{\today}
\maketitle

Bulk-boundary correspondence is a defining feature of topological state of matters. The bulk bands of topological matters are characterized by topological invariants and there are nontrivial boundary states corresponding to the nontrivial bulk bands\cite{PhysRevLett.95.146802,PhysRevLett.98.106803,PhysRevB.78.195424}. Wilson-loop serves as a bridge between the bulk property and the boundary property: the bulk topological invariant and the boundary state property can both be derived from the Wilson-loop\cite{PhysRevB.84.075119,PhysRevB.83.235401,PhysRevB.83.035108,PhysRevLett.107.036601}.

The phases of eigenvalues the Wilson-loop along one direction are proportional to the centers of the maximumly localized Wannier functions\cite{PhysRevB.56.12847}. In 2D or 3D  sytems the Wannier functions are hybrid\cite{PhysRevB.83.035108}, that is, they are localized in one direction and extended in other directions.  By studying the Wannier centers as function of $\mathbf{k}_{||}$ we can get the topological information of the bulk bands\cite{PhysRevB.84.075119,PhysRevB.83.235401,PhysRevB.83.035108,PhysRevB.89.155114}, where $\mathbf{k}_{||}$ is the wave vector in the direction that the Wannier functions are extended. In this letter the Wannier centers as functions of $\mathbf{k}_{||}$ are referred to as Wilson-loop spectrums.

On the other hand, spectrum of Wilson-loop can be continuously deformed to boundary state spectrum\cite{PhysRevLett.107.036601}. So the bulk-boundary correspondence can be established by studying the topology of Wilson-loop spectrum.

Despite of its great versatility in characterizing topological property of matters, up to now the Wilson-loop exists only in theory. That is, the Wilson-loops are only used as auxiliary tools either for computing the topological invariant\cite{PhysRevB.84.075119,PhysRevB.83.235401,PhysRevB.83.035108} or for characterizing the topology of boundary states\cite{PhysRevLett.107.036601}. No attempt has been made to measure the topology of Wilson-loop.

For all the applications of Wilson-loop, the knowledge of the topology of Wilson-loop spectrum is enough\cite{PhysRevB.84.075119,PhysRevB.83.235401,PhysRevB.83.035108,PhysRevLett.107.036601}. So we can obtain the topology of Wilson-loop by measuring something that is topologically equivalent to the Wilson-loop spectrum. In ref.\cite{PhysRevLett.107.036601} it is shown that the boundary-state spectrum is a continuous deformation of Wilson-loop spectrum if the bulk system is gapped. Thus it is natural to obtain the topology of Wilson-loop spectrum by measuring the boundary-state spectrum. However, as we show in supplementary  material\cite{see}, the boundary-state spectrum is only a partial deformation of the Wilson-loop spectrum in general and this partial information from the boundary-state spectrum is sometime misleading. In the topological semimetals, where the bulk system is not gapped, the surface state spectrum is not a partial deformation of Wilson-loop spectrum at all.

Another defect of obtaining topological information of Wilson-loop from boundary state spectrum is that boundary state spectrum can not reflect the periodic feature of Wilson-loop spectrum. Wilson-loop spectrum is defined modulo a lattice constant. So it is periodic in the direction that the Wannier functions are localized. When Wilson-loop spectrum deforms to boundary state spectrum this periodic feature disappears. So we need a systematic way to measure the topology of Wilson-loop spectrum.

\begin{figure}
\includegraphics[width=8.6cm]{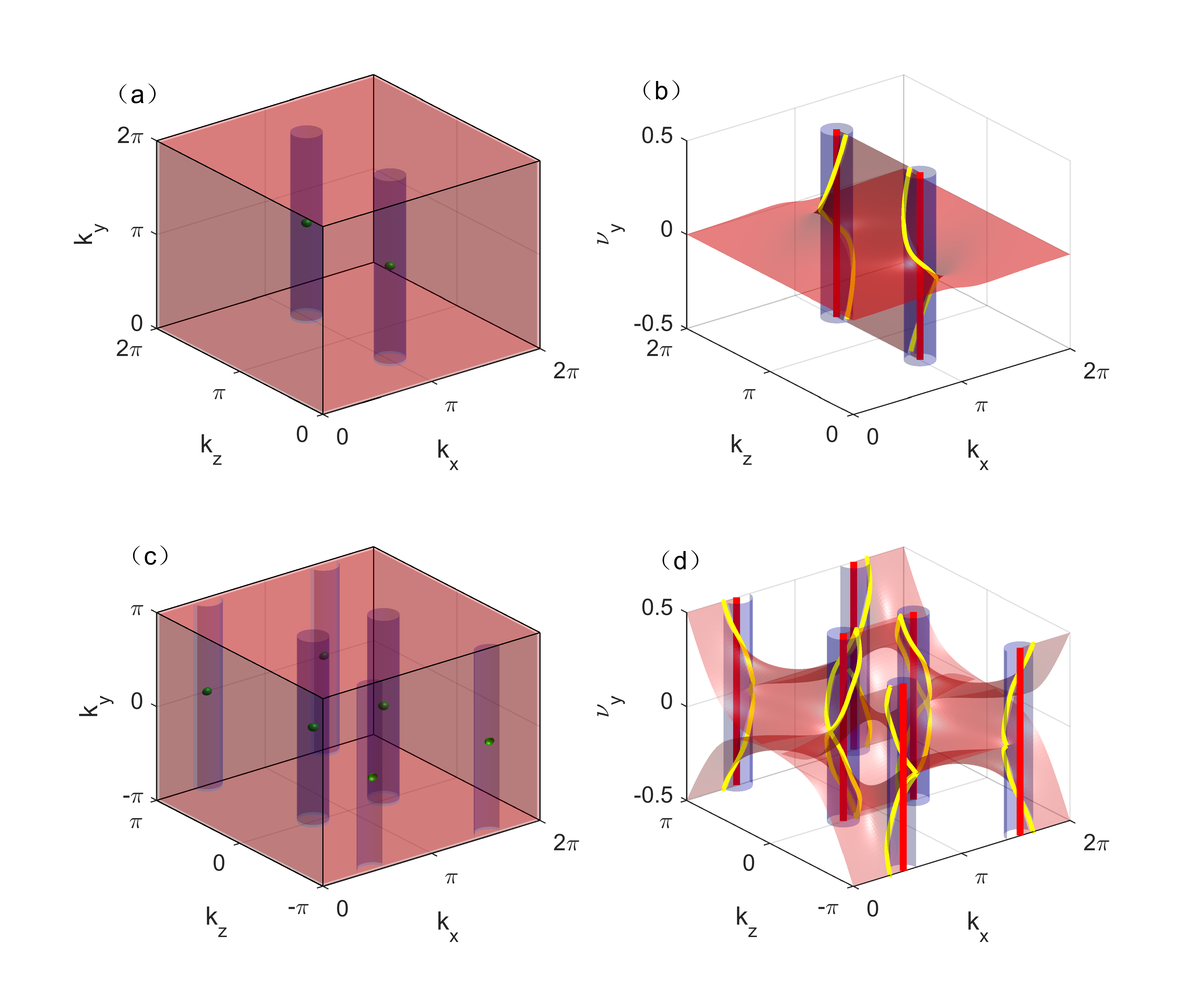}
\caption{(a) and (c) Brillouin zone of Weyl semimetal and Dirac semimetal.  The spheres in the cylinders are Weyl points and Dirac points.
(b) and (d) Behavior of the Wilson-loop spectrums of topological semimetals near the Weyl and Dirac points. The intersections of the Wilson-loop spectrums and the cylinders are highlighted by yellow curves.
}

	\label{fig1}
\end{figure}

Our work is inspired by the recent interests in surface-state Fermi arc of topological matters, especially, of topological Weyl and Dirac semimetals\cite{Xu294,Xu613,yang2017direct,Noh2017Experimental,li2018weyl,PhysRevApplied.10.014017,PhysRevLett.122.104302}. The surface-state Fermi arcs have been observed in both electronic\cite{Xu294,Xu613} and artificial periodic structures, such as phononic crystals and photonic crystals\cite{yang2017direct,Noh2017Experimental,li2018weyl,PhysRevApplied.10.014017,PhysRevLett.122.104302}.

The spectrum of surface states is sensitive to  boundary conditions\cite{HE20101976}. So the surface-state Fermi arc is also boundary-dependent\cite{he2018topological}. This boundary-dependence of Fermi arc has been used to realize the topological negative refraction of surface acoustic waves in a Weyl phononic crystal\cite{he2018topological}. When the boundary of the system varies continuously with a parameter, the Fermi arc changes with the parameter and thus generates a surface.

If the system is half infinite, when one layer of unit cells are removed from the system in one continuous process, the process can be considered as periodic.
In supplemental material\cite{see} we show the evolution of Fermi arc in such periodic process parameterized by $\phi_d$ generates a surface that is topologically equivalent to the Wilson-loop spectrum. We call this surface the ``boundary Fermi surface'' in following discussions.

The topological equivalence between Wilson-loop spectrum and ``boundary Fermi surface'' is based on the fact that the bulk system is gapped. However, in Weyl or Dirac semimetals the bulk gap closes at the Weyl or Dirac points. Here we show the topological equivalence is not affected by the presence of Weyl or Dirac points.

To establish the topological equivalence we divided the Brillouin zone of the topological semimetal into two parts. One part consists of the open cylinders that include the Weyl or Dirac points as is show in Fig.1 (a) and (c). The rest of Brillouin zone constitutes the second part. Clearly in the second part the bulk energy spectrum is gapped. So topological equivalence can be established in this part by using the arguments in supplementary material\cite{see}. Then we let the radius of the cylinders approaches zero. In this process the topological equivalence does not change because it does not depends on the size of bulk band gap. In this way we prove the topological equivalence in the whole Brillouin zone.

One interesting phenomenon is that the Wilson-loop spectrum approaches to vertical lines at projection points the Weyl or Dirac points. For example, if the cylinder encloses a Weyl point with monopole charge $q$, the integral of the
Berry flux over the surface of the cylinder is $q2\pi$. If we consider the surface of one cylinder as a 2D system, the system will have a Chern number $q$. It is well know the Wilson-loop spectrum of a system with nonzero Chern number is gapless. So when the radiuses of the cylinders approach to zero the Wilon-loop spectrum will approach to vertical lines as is shown in Fig.1 (b) and (d). To the best of our knowledge, no proposal have been made to use this topological feature to characterize the topological semimetals.

\begin{figure}
\includegraphics[width=8.6cm]{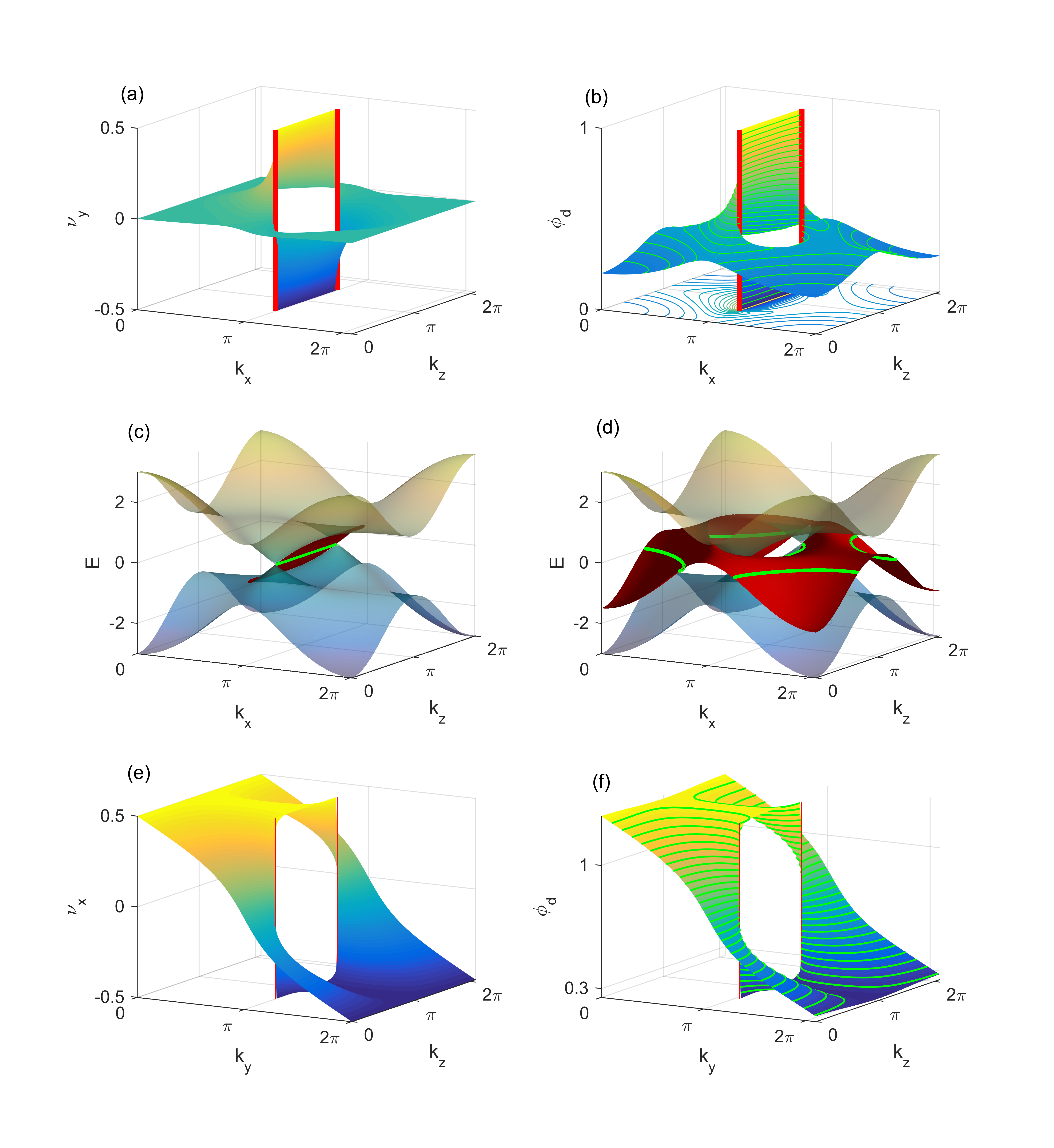}
		    \caption{
(a)Wilson-loop spectrum along the $y$ direction of the Weyl semitmetal of Model (1).
(b)``Boundary Fermi surface'' of the same weyl smeimetal. It is generated by the Fermi arcs of the surface states on the $XZ$ surface at Fermi energy $E_F=0$.
(c) and (d) Surface state spectrum of the this model with two different boundary conditions. $\phi_d=0$ in (c) and $\phi_d=0.25$ in (d). The red surfaces are the surface state spectrums.  The yellow curves are Fermi arc of surface states at $E_F=0$. The semi-transparent surfaces correspond to the bottom of upper bulk bands and top of the lower bulk bands. They form Dirac cones at the neighbourhood of the project points of the Wyel points. (e) and (f) Wilson-loop spectrum and ``boundary Fermi surface'' of the continuous model (2). $\hbar=m=V_1=V_2/2=1$ in the calculation. In (f) the Fermi energy equals to the energy at the Weyl points.
}
	\label{fig1}
\end{figure}

To illustrate our conclusion we use the simplest tight-binding model for Weyl semimetal
\begin{equation}
\begin{split}
H=&\sin k_x \sigma_x+\sin k_y \sigma_y\\
&+ (M_0+\cos k_x+\cos k_y+\cos k_z)\sigma_z,
\end{split}
\end{equation}
 where $\sigma_i$, $i=x,y,z$, are the Pauli matrices. There are two Weyl points at $\mathbf{k}=(\pi,\pi,\pi\pm\pi/2)$ when $M_0=2$. In Fig.2 (a) and (b) we show the topological equivalence between Wilson-loop spectrum and ``boundary Fermi surface'' of the Weyl semimetal. Especially, we show the ``boundary Fermi surface'' approaches to vertical lines at the projection points of the Weyl points.
  In comparison, the surface state spectrums in Fig.2 (c) and (d) fail to reflect this topological feature of the Wilson-loop spectrum. It is easily seen from Fig.2 (a), (c) and (d) the surface state spectrums of the Weyl semimetal are not partial deformations of the Wilson-loop spectrum.

\begin{figure}
\includegraphics[width=8.6cm]{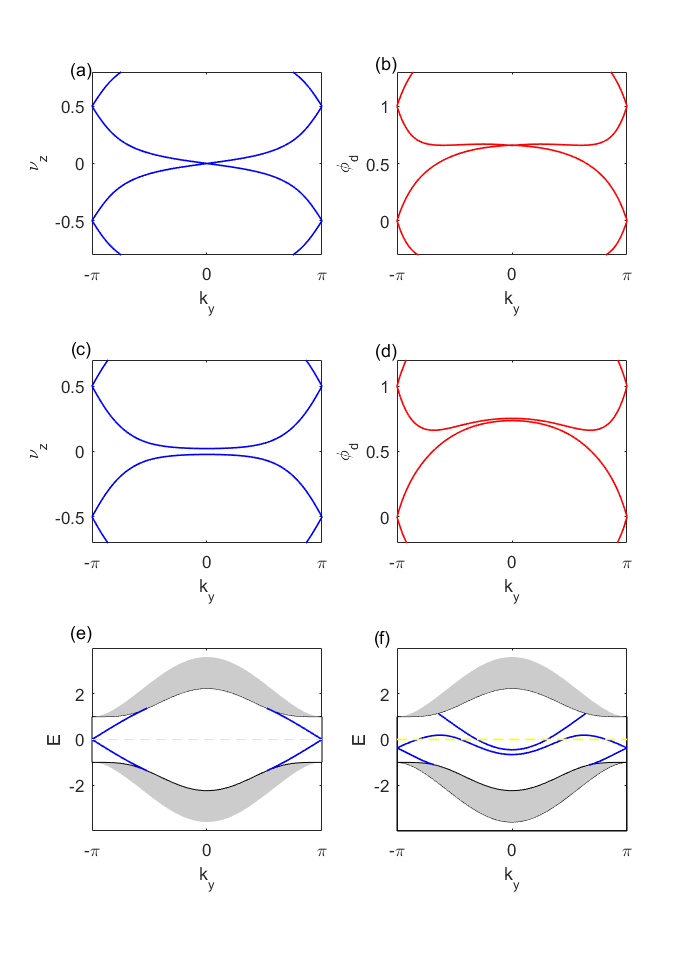}
		    \caption{
(a) Wilson-loop spectrum along the $z$ direction of the 2D TI. $M=2$, $t=\Delta_1=1$ and $\Delta_2=0$ in the calculations.
(b) ``Boundary Fermi surface'' of the 2D TI.
(c) Wilson-loop spectrum along the $z$ direction of a 2D trivial insulator which can exhibit gapless edge states with some boundary conditions.
(d) ``Boundary Fermi surface'' of the trivial insulator.
(e) and (f), Edge state spectrums of the trivial insulator with different boundary conditions. In (e)((f)) $\phi_d=0$($\phi_d=0.75$) and the edge state spectrum is gapless(gapped).
}
	\label{fig1}
\end{figure}

It is technically challenging to continuously remove one layer of unit cells from the electronic system without breaking the translational symmetry in the parallel direction. However,  in classical phononic crystals and photonic crystals, the unit cells are of macro scale and the boundary of the system can be easily controlled.

We can remove one layer of unit cells from these artificial structures by cutting the unit cells slice by slice. The surface-state Fermi arc of the system is measured when each slice is cut off. Such a process can be considered as an approximation of the continuous process we propose. The thickness(in unit of lattice constant) of the layer that is cut off can be chosen as the parameter $\phi_d$. So we can use the exiting experimental setup for measuring surface-state Fermi arc to detect the topology of Wilson-loop spectrum of the artificial structures. The only extra effort is to cut the unit cells, which is particularly easy for phononic crystals because the unit cells are fabricated by 3D printing\cite{li2018weyl,PhysRevApplied.10.014017,PhysRevLett.122.104302}.
\begin{figure}
\includegraphics[width=8.6cm]{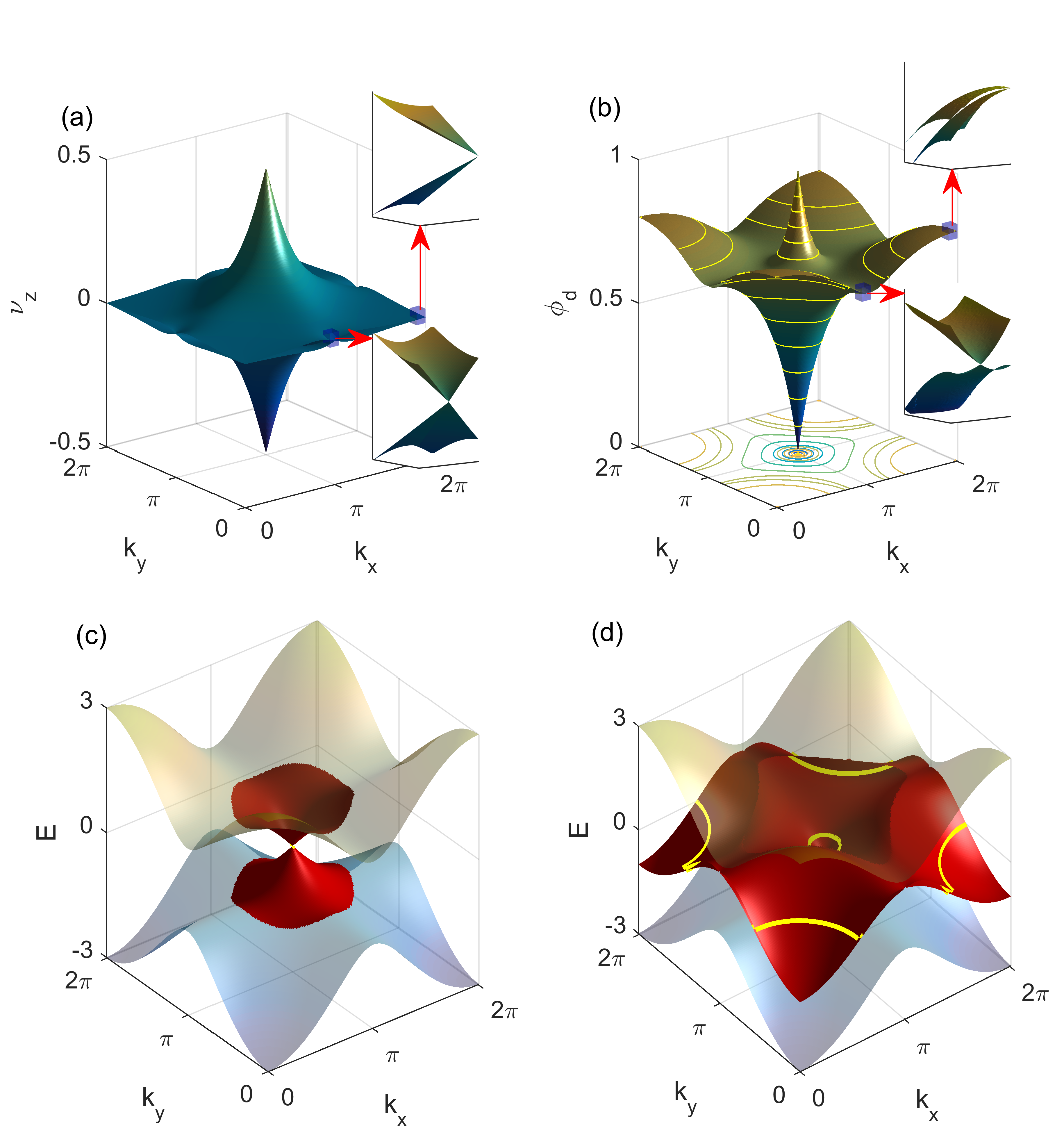}
		    \caption{
(a) Wilson-loop spectrum along the $z$ direction of the 3D TI. The insets indicated the gaplessness of the Wilson-loop spectrum. The parameters take the same value as in the 2D TI.
(b) ``Boundary Fermi surface'' of the same TI. The insets show it reflects the gaplessness of Wilson-loop spectrum.
(c) and (d) Surface state spectrum of the same bulk TI with two different boundaries. $\phi_d=0$ in (c) and $\phi_d=0.75$ in (d). The red surfaces are the surface state spectrums. The yellow cures are Fermi arcs of surface states at $E_F=0$. In (c) the Fermi arc reduces to a point. The semi-transparent surfaces also correspond to the bottom of upper bulk bands and top of the lower bulk bands. They are gapped in 3D TI.
}
	\label{fig3}
\end{figure}

Since we predict the topology of Wilson-loop spectrum can be observed in phononic crystal and since the rigid boundary conditions are used when Fermi arc is measured in phononic crystal\cite{li2018weyl,PhysRevApplied.10.014017,PhysRevLett.122.104302}, it is more convincing to use a continuous model and rigid boundary condition to show the topological equivalence between Wilson-loop spectrum and ``boundary Fermi surface''. We use a 1D quantum model with two parameters to simulate a 3D Wyel semimetal. The Hamiltonian of the semimetal can be expressed as
\begin{equation}
\begin{split}
H=& -\frac{\hbar^2}{2m}\frac{d^2}{dx^2}+V_1\cos(2\pi x/a)\\
   &+V_2\sin^2(k_z)\cos(2\pi x/a-k_y),
\end{split}
\end{equation}
 where a is lattice constant, $V_1$ and $V_2$ are constants. $k_y$ and $k_z$ are two parameters to map the system to a 3D system. When $|V_2|>|V_1|>0$ it represents a Weyl semimetal. The rigid boundary condition can be realized by choosing $\psi(\phi_d a)=0$. If the system is on the right side of $x=\phi_d a$, one layer of unit cells will be continuously removed from the system when $\phi_d$ continuously increases by one. In Fig.2 (e) and (f) we show the topological equivalence between Wilson-loop spectrum and ``boundary Fermi surface'' of this model.

The topology of Wilson-loop spectrums of 2D and 3D topological insulators(TIs) can be similarly measured. The models for 2D and 3D TIs used in this letter are all adapted from the model for chiral high order topological insulator(HOTI) in ref.\onlinecite{schindler2018higher-order}
\begin{equation}
\begin{split}
H(\mathbf{k})=&(M+t\sum_i\cos k_i)\tau_z\sigma_0+\Delta_1 \sum_i\sin k_i \tau_x\sigma_i\\
          & +\Delta_2(\cos k_x-\cos k_y)\tau_y\sigma_0,
\end{split}
\end{equation}
where $\sigma_i$ and $\tau_i$,$i=x,y,z$, are Pauli matrices acting on spin and orbital degree of freedoms, respectively. When $\Delta_2=0$ the time-reversal symmetry breaking term disappears and it is a model for strong 3D TI. The model for 2D TI can be obtained by setting $k_x=\pi$ with $\Delta_2=0$.

In Fig.3 and Fig.4 we show the topological equivalence between ``boundary Fermi surfaces'' and Wilson-loop spectrums of the TIs.
In the 2D systems the ``surface-state Fermi arcs'' are points in 1D Brillouin zone. The evolution of these points generate curves that are topologically equivalent to the Wilson-loop spectrum.

If the bulk bands are gapped, the spectrums of edge or surface states are partial deformation of Wilson-loop spectrums as is shown in Fig.3 (e) and (f) and Fig.4 (c) and (d). However, this partial information from boundary-state spectrum can be misleading as is indicated in Fig.2 (c) and (e). In Fig.2 (c) we show the Wilson-loop spectrum of a 2D trivial insulator. The model for this trivial insulator is constructed by setting $k_x=\pi$ with $\Delta_2=1$ in (3). We can see the Wilson-loop spectrum of this model is gapped. However,  in Fig.3 (e) and (f) we show that the edge state spectrums can be both gapless and gapped depending on the choice of boundary conditions. In contrast, the ``boundary Fermi surface'' in Fig.3 (d) contains all the topological information of the Wilson-loop spectrum.

The topology of Wilson-loop spectrum of HOTIs\cite{Benalcazar61,PhysRevB.96.245115,schindler2018higher-order,imhof2018topolectrical-circuit,serragarcia2018observation,zhang2019second-order,PhysRevLett.123.073601} can also be measured by observing the evolution of surface-state Fermi arc. However, the Wilson-loop spectrums of HOTIs are gapped as that of trivial topological insulators. So Wilson-loop spectrum can not be used to characterize the topology of HOTIs. The HOTIs distinguish themselves from trivial insulators by the fact that the Wannier bands carry nontrivial topological invariant. The topological invariant of Wannier band can be obtained from the topology of nested Wilson-loop spectrum. The topology of nested Wilson-loop spectrum can also be observed if we measure the boundary wave functions in the periodic process measuring the ``boundary Fermi surfaces''\cite{to}.

\bibliographystyle{apsrev4-1}

\begin{thebibliography}{27}
\bibitem [{to()}]{to}%
  \BibitemOpen
  \href@noop {} {\bibinfo  {journal} {To be disscussed later}\ }\BibitemShut
  {NoStop}%
\end{thebibliography}

\end{document}